\documentclass[conference]{IEEEtran}
\usepackage{cite}
\usepackage{amsmath,amssymb,amsfonts}
\usepackage{algorithmic}
\usepackage{graphicx}
\usepackage{textcomp}
\usepackage{xcolor}
\graphicspath{{Figures/}}
\DeclareGraphicsExtensions{.pdf,.eps}
\usepackage[outdir=Figures/]{epstopdf}
\usepackage{grffile}
\usepackage{xspace}
\PassOptionsToPackage{nobreak}{cite}
\usepackage{flushend}
\usepackage{comment}
\usepackage{color}
\usepackage{url}
\usepackage{hyperref}
\hypersetup{
        colorlinks=true,
        allcolors=black,
        bookmarksnumbered=true,
        bookmarkstype=toc,
        bookmarksopen=true,
        pdftitle={Exploring Optimal Deep Learning Models for Image-based Malware Variant Classification},
        pdfauthor={},
        pdfstartview={FitH -32768}
}

\usepackage{tikz}
\usepackage{multirow}
\usepackage{pdfpages}   

\newcommand{\doi}{DOI:10.1109/COMPSAC54236.2022.00128}
\newcommand{\doiurl}{https://doi.org/\doi}

\usepackage{setspace}
\makeatletter
\def\@IEEEpubidpullup{9\baselineskip}
\makeatother

\IEEEoverridecommandlockouts
\IEEEpubid{
\parbox{\columnwidth}{
\vspace{-10\baselineskip}
\noindent\rule{\columnwidth}{0.5pt}
\copyright \space 2022 IEEE. Personal use of this material is permitted. Permission from IEEE must be obtained for all other uses, in any current or future media, including reprinting/republishing this material for advertising or promotional purposes, creating new collective works, for resale or redistribution to servers or lists, or reuse of any copyrighted component of this work in other works.
\vspace{0.5ex}\\
Rikima Mitsuhashi and Takahiro Shinagawa. Exploring Optimal Deep Learning Models for Image-based Malware Variant Classification. In Proceedings of the IEEE 46th Annual Computers, Software, and Applications Conference (IEEE COMPSAC 2022), pp. 779--781, Jun 2022.\\
\href{\doiurl}{\doi}
}
\hspace{0.9\columnsep}\makebox[\columnwidth]{\hfill}}
\IEEEpubidadjcol

\usepackage{nth}
\usepackage{booktabs}    

\begin{document}
\title{
\begin{spacing}{1}
\normalsize \it
This is the accepted version of the paper published in \href{https://ieeecompsac.computer.org/2022/}{IEEE COMPSAC 2022},\\
publicly downloadable from \href{http://www.os.ecc.u-tokyo.ac.jp}{the Shinagawa Laboratory's website}.\\
The published version is available at \href{\doiurl}{\doi}.
\end{spacing}
\vspace{3ex}
Exploring Optimal Deep Learning Models for Image-based Malware Variant Classification
}

\author{
\IEEEauthorblockN{Rikima Mitsuhashi}
\IEEEauthorblockA{\textit{The University of Tokyo} \\
Tokyo, Japan \\
mitsuhashi@os.ecc.u-tokyo.ac.jp}
\and
\IEEEauthorblockN{Takahiro Shinagawa}
\IEEEauthorblockA{\textit{The University of Tokyo} \\
Tokyo, Japan \\
shina@ecc.u-tokyo.ac.jp}
}

\maketitle

\begin{abstract}
Analyzing a huge amount of malware is a major burden for security analysts.
Since emerging malware is often a variant of existing malware, automatically classifying malware into known families greatly reduces a part of their burden.
Image-based malware classification with deep learning is an attractive approach for its simplicity, versatility, and affinity with the latest technologies.
However, the impact of differences in deep learning models and the degree of transfer learning on the classification accuracy of malware variants has not been fully studied.
In this paper, we conducted an exhaustive survey of deep learning models using 24 ImageNet pre-trained models and five fine-tuning parameters, totaling 120 combinations, on two platforms.
As a result, we found that the highest classification accuracy was obtained by fine-tuning one of the latest deep learning models with a relatively low degree of transfer learning, and we achieved the highest classification accuracy ever in cross-validation on the Malimg and Drebin datasets.
We also confirmed that this trend holds true for the recent malware variants using the VirusTotal 2020 Windows and Android datasets.
The experimental results suggest that it is effective to periodically explore optimal deep learning models with the latest models and malware datasets by gradually reducing the degree of transfer learning from half.
\end{abstract}


\section{Introduction}
\label{sec:introduction}

In the field of cybersecurity, security analysts are struggling with a huge amount of malware.
Since recent malware is more complex and sophisticated than ever before, security analysts must carefully analyze its structure and behavior based on their advanced knowledge and insights.
Unfortunately, recent malware often exploits new vulnerabilities and obfuscation techniques~\cite{McAfee2021}, leaving security analysts with the burden of studying the latest technologies and performing manual analysis.
Moreover, even novice users can generate new malware using easily available tools, such as malware generators that run on smartphones~\cite{Symantec} and malware as a service (MaaS)~\cite{Kaspersky}, which is spurring the increase of malware.

One way to reduce the burden of malware analysis is automatic malware classification.
Since attackers tend to create many \emph{malware variants}  to save time and effort in creating new ones, automatically classifying malware into known families helps security analysts to leverage their past knowledge and experience.
Machine learning (ML) is a promising approach to automatic malware classification, and many ML-based methods have been proposed~\cite{UCCI2019123,SINGH2021101861}.
However, most ML-based methods require security experts to manually define malware features, which is time-consuming, task-specific, error-prone, and highly dependent on individual experience and knowledge~\cite{10.1145/3417978}.
Deep learning overcomes this problem by using a deep hierarchical data model that automatically extracts nonlinear features at various levels of abstraction without manual efforts~\cite{10.1145/3417978}. 

A convolutional neural network (CNN) is a popular algorithm in deep learning because of the availability of huge datasets, many properly pre-trained models, and support for transfer learning.
Many studies tried to extract malware features suitable for CNNs, such as opcode sequences, control flow graphs, and API calls~\cite{10.1145/3417978}.
However, manual feature extraction requires expert knowledge and skills, and its effectiveness can vary from platform to platform and over time.
Malware images~\cite{10.1145/2016904.2016908}, which are generated by converting malware binaries directly into grayscale images, is a simple yet versatile approach that is expert-knowledge-free, applicable to any platform, and highly compatible with image classification methods.
In addition, since CNNs have been actively researched and developed for image recognition, the latest technologies can be easily incorporated.
However, previous studies on applying CNNs to malware image classification~\cite{8330042,8260773,10.1145/3320326.3320333,8328749,8763852,10.1145/3194452.3194459,forse18,forse19,VASAN2020107138} have limited variations of models and transfer learning parameters, and their impact on classification accuracy has not been fully explored.

In this paper, we conducted an exhaustive study on the accuracy of image-based malware classification with a number of CNN models.
We used CNN models pre-trained with ImageNet~\cite{5206848} provided by Keras~\cite{Keras}. 
ImageNet is an image database with supervised labels consisting of more than 14 million color photographs of objects that exist in nature, such as plants, terrain, sports, animals, and so on.
Thus, ImageNet pre-trained models would have some of the visual recognition capabilities of humans.
However, it is unclear how effective the ability to distinguish objects in nature is in classifying artificially created malware images.
Therefore, we investigated the effect of ImageNet pre-training on malware classification accuracy with different degrees of transfer learning.
Specifically, we prepared 24 pre-trained models and five fine-tuning parameters for each model, 120 combinations in total.
To the best of our knowledge, this is the first study to investigate the effects of different CNN pre-training models and fine-tuning parameters on malware classification accuracy in such a wide variety of combinations.

We conducted our study in two steps.
First, to identify the optimal CNN models to classify malware images, we used two pre-labeled datasets: Malimg~\cite{Malimg} for Windows and Drebin~\cite{Drebin} for Android.
As a result, we found that the EfficientNetB4 model fine-tuned with no frozen parameters, i.e., forgetting all ImageNet pre-trained knowledge, had the highest classification accuracy in Mailimg, obtaining 98.96\% with cross-validation. 
For the top 20 classes of Drebin, the EfficientNetB4 model fine-tuned with 1/4 of the pre-trained parameters frozen achieved the highest classification accuracy of 91.03\% with cross-validation.
To the best of our knowledge, these are the highest classification accuracy of Malimg and Drebin with cross-validation reported so far.

Next, to examine the classification accuracy against recent real malware, we obtained the VirusTotal~\cite{VirusTotal} May 2020 datasets and trained several models that had relatively high classification accuracy in Malimg and Drebin with the datasets.
As a result, for Windows malware, EfficientNetB5 with 1/2 of the pre-trained parameters frozen achieved 98.78\% accuracy, and for Android malware, EfficientNetB3-B5 with no frozen parameters achieved 100\% accuracy. 

From the results of these experiments, we found that the highest malware variant classification accuracy tends to be obtained by using the latest deep learning models with a relatively low degree of transfer learning, i.e., freezing the fine-tuning parameters from 0\% to 50\%.
This suggests that the ability to recognize natural objects is effective to some extent, but learning more features of malware images is also important in classifying malware variants.
Since it is not easy to identify in advance the optimal parameters for each malware dataset, a realistic approach is to periodically explore the optimal degree of transfer learning by fine-tuning the latest model with the latest malware data while gradually lowering the freezing parameters from 50\%.

The contributions of this paper are as follows:
\begin{itemize}
\item We conducted an exhaustive study on the impact of different models and fine-tuning parameters of CNNs, 120 combinations in total, on the classification accuracy of malware variants.
\item We found that the optimal model for malware classification of Malimg and Drebin was EfficientNetB4 with pre-trained parameters unfrozen and 25\% frozen, respectively, and achieved the highest classification accuracy to date with cross-validation.
\item We confirmed that fine-tuning the latest models with lower frozen parameters is effective for classifying malware variants in VirusTotal; EfficientNetB5 with 50\% frozen for Windows and EfficientNetB3-B5 with no frozen for Android achieved the highest accuracy.
\end{itemize}

The rest of this paper is organized as follows:
\autoref{sec:related} presents related work.
\autoref{sec:background} explains background knowledge on CNNs and fine-tuning.
\autoref{sec:method} describes our experimental method to explore the optimal models for malware variant classification.
\autoref{sec:evaluation} shows the experimental results and
\autoref{sec:conclusion} concludes the paper.

\section{Related Work}
\label{sec:related}

\begin{figure}[t]
	\centering
	\includegraphics[width=\linewidth,clip]{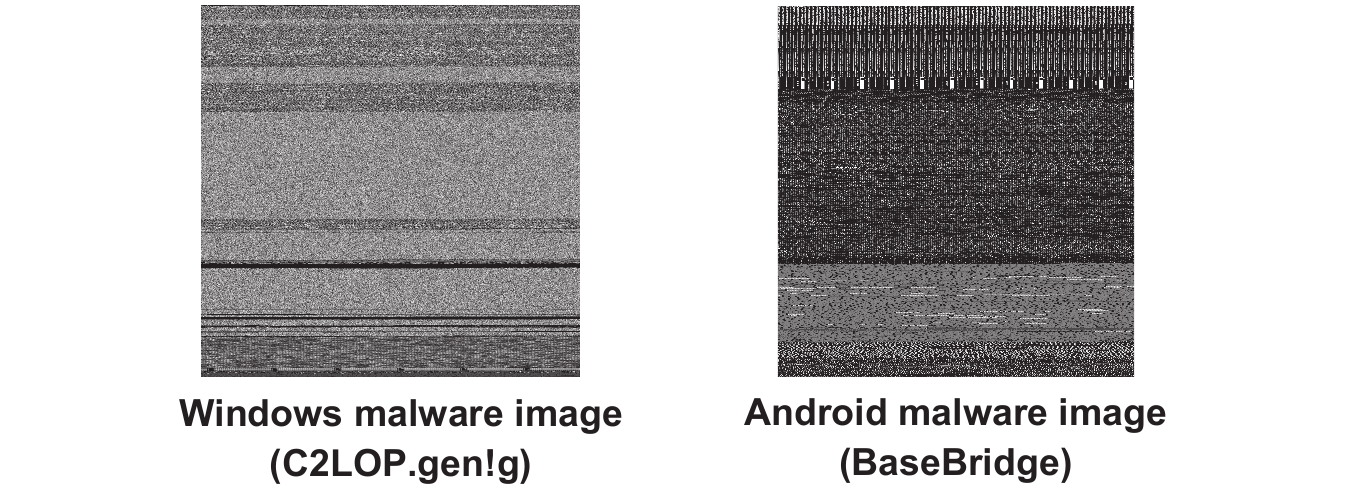}
	\caption{Example of Windows and Android malware images.}
	\label{fig_malwareimage}
\end{figure}

Malware image is a concept whereby malware binaries are converted into visual images to extract malware features~\cite{10.1145/2016904.2016908}. Malware image has the advantage that features can be easily extracted without expert knowledge and that they can be applied to different platforms, such as Windows and Android. \autoref{fig_malwareimage} shows the example of malware images.

Many previous studies have attempted to improve the accuracy of image-based classification using Malimg as a standard dataset.
Nataraj et al.~\cite{10.1145/2016904.2016908} used GIST and k-nearest neighbors for classification and obtained an accuracy of 97.18\% with 10-fold cross-validation.
Kosmidis and Kalloniatis~\cite{10.1145/3139367.3139400} used several ML techniques such as decision tree, nearest centroid, stochastic gradient, perceptron, multilayer perceptron, and Random Forest. The best result was Random Forest with 91.6\% average accuracy.
Cui et al.~\cite{8330042} used a CNN and addressed the data imbalance among different malware families.
The classification accuracy of cross-validation was 94.5\%.
Rezende et al.~\cite{8260773} prepared a ResNet50 model with all convolutional layer parameters transferred from a model trained with ImageNet and achieved the classification accuracy of 98.62\% with cross-validation.
Mourtaji et al.~\cite{10.1145/3320326.3320333} reported the VGG16 model with an accuracy of 97.02\% using the hold-out method.
Kalash et al.~\cite {8328749} proposed M-CNN based on VGG16 pre-trained by ImageNet and achieved an accuracy of 98.52\% with hold-out validation.
Lo et al.~\cite{8763852} used the Xception model and its classification accuracy reached 99.03\% using the hold-out method.

Several studies tried to manipulate malware images or machine learning models to improve classification accuracy.
Vasan et al.~\cite{VASAN2020107138} converted malware binaries into color images and had a classification accuracy of 98.82\%, although their evaluation was not cross-validation.
Su et al.~\cite{8377943} used a two-layer shallow convolutional neural network and achieved 94.0\% accuracy in classifying goodware and DDoS malware with 5-fold hold-out validation.
Vasan et al.~\cite{VASAN2020101748} introduced an ensemble method that used five models and had an accuracy of 99.50\% with hold-out validation.
Although their ensemble method used an argmax function for finding the model with the highest predicted probability, they did not evaluate which of the five models was the most important for improving the classification accuracy. 

For image-based classification of Android malware, Singh et al. ~\cite{9461194} classified the top 20 Drebin dataset using images generated from manifest and certificate of android application. 
As a result, they obtained a classification accuracy of 93.24\% using the Feature Fusion-SVM model.

Although all the previous studies have shown good classification accuracy, there is still room for improvement.
Even a 0.1\% improvement in accuracy has a non-negligible impact when there are tens of thousands of malware variants.
Moreover, few studies investigated the impact of different models and fine-tuning parameters on classification accuracy.

\begin{table*}[t]
	\caption{The numbers and names of frozen layers in the Keras implementation for our 24 models.}
	\label{tab_targetlayer}
	\centering
	\scalebox{0.95}{
	\begin{tabular}{l|lllll} 
		\toprule
		& Frozen all  & Frozen 3/4  & Frozen 1/2 & Frozen 1/4 & Frozen none \\
		\midrule
				DenseNet121       & 427 (relu)  & 341 (conv5\_block4\_concat)  & 274 (conv4\_block19\_concat) & 162 (conv4\_block3\_concat)  & 1 (input\_1) \\
DenseNet169       & 595 (relu)  & 495 (conv5\_block18\_concat) & 365 (conv4\_block32\_concat) & 253 (conv4\_block16\_concat) & 1 (input\_1) \\
DenseNet201       & 707 (relu)  & 572 (conv5\_block13\_concat) & 449 (conv4\_block44\_concat) & 316 (conv4\_block25\_concat) & 1 (input\_1) \\
EfficientNetB0    & 230 (top\_activation) & 214 (block6d\_add) & 184 (block6b\_add) & 156 (block5c\_add) & 1 (input\_1) \\
EfficientNetB1    & 332 (top\_activation) & 314 (block7a\_project\_bn) & 286 (block6d\_add) & 241 (block6a\_project\_bn) & 1 (input\_1) \\
EfficientNetB2    & 332 (top\_activation) & 314 (block7a\_project\_bn) & 286 (block6d\_add) & 241 (block6a\_project\_bn) & 1 (input\_1) \\
EfficientNetB3    & 377 (top\_activation) & 359 (block7a\_project\_bn) & 316 (block6d\_add) & 271 (block6a\_project\_bn) & 1 (input\_1) \\
EfficientNetB4    & 467 (top\_activation) & 449 (block7a\_project\_bn) & 391 (block6e\_add) & 331 (block6a\_project\_bn) & 1 (input\_1) \\
EfficientNetB5    & 569 (top\_activation) & 551 (block7b\_add) & 493 (block6g\_add) & 418 (block6b\_add) & 1 (input\_1) \\
InceptionResNetV2 & 780 (conv\_7b\_ac) & 698 (block8\_5\_ac) & 618 (mixed\_7a) & 419 (block17\_9\_ac) & 1 (input\_1) \\
InceptionV3       & 311 (mixed10) & 280 (mixed9) & 249 (mixed8) & 165 (mixed5) & 1 (input\_1) \\
MobileNet         & 87  (conv\_pw\_13\_relu) & 84  (conv\_dw\_13\_relu) & 74  (conv\_pw\_11\_relu) & 56 (conv\_pw\_8\_relu) & 1 (input\_1) \\
MobileNetV2       & 155 (out\_relu) & 144 (block\_15\_add) & 135 (block\_14\_add) & 117 (block\_12\_add) & 1 (input\_1) \\
NASNetLarge       & 1039 (activation\_259) & 948 (normal\_concat\_16) & 858 (normal\_concat\_14) & 711 (normal\_concat\_12) & 1 (input\_1) \\
NASNetMobile      & 769 (activation\_187) & 723 (normal\_concat\_11) & 633 (normal\_concat\_9) & 531 (normal\_concat\_8) & 1 (input\_1) \\
ResNet101         & 345 (conv5\_block3\_out) & 325 (conv5\_block1\_out)  & 253 (conv4\_block17\_out) & 163 (conv4\_block8\_out)  & 1 (input\_1) \\
ResNet152         & 515 (conv5\_block3\_out) & 483 (conv4\_block36\_out) & 353 (conv4\_block23\_out) & 223 (conv4\_block10\_out) & 1 (input\_1) \\
ResNet50          & 175 (conv5\_block3\_out) & 165 (conv5\_block2\_out)  & 155 (conv5\_block1\_out)  & 123 (conv4\_block4\_out)  & 1 (input\_1) \\
ResNet101V2       & 377 (post\_relu) & 353 (conv5\_block1\_out)  & 274 (conv4\_block17\_out) & 175 (conv4\_block8\_out)  & 1 (input\_1) \\
ResNet152V2       & 564 (post\_relu) & 528 (conv4\_block36\_out) & 384 (conv4\_block23\_out) & 241 (conv4\_block10\_out) & 1 (input\_1) \\
ResNet50V2        & 190 (post\_relu) & 177 (conv5\_block2\_out)  & 166 (conv5\_block1\_out)  & 131 (conv4\_block4\_out)  & 1 (input\_1) \\
VGG16             & 19  (block5\_pool) & 17 (block5\_conv2)  & 15 (block4\_pool)  & 12 (block4\_conv1)  & 1 (input\_1) \\
VGG19             & 22  (block5\_pool) & 19 (block5\_conv2)  & 17 (block4\_pool)  & 14 (block4\_conv2)  & 1 (input\_1) \\
Xception          & 132 (block14\_sepconv2\_act) & 126 (add\_11)  & 96 (add\_8)  & 66 (add\_5)  & 1 (input\_1) \\
		\bottomrule
	\end{tabular}
	}
\end{table*}

\section{Background}
\label{sec:background}

In this section, we briefly explain the basics of CNNs and fine-tuning, which are necessary to understand the experiments described later.

\subsection{CNN}

A CNN is a class of deep neural network (DNN), which is an evolution of an artificial neural network (ANN)~\cite{10.1145/3234150}.

ANNs are designed to mimic the neural structure of human brains, where many neurons are connected by synapses.
Information is transmitted between neurons by electrical signals, and the amount of information transmitted is determined by the strength of the synaptic connections.
A neuron that receives signals from multiple neurons produces a single output when the sum of the signals exceeds a certain threshold.
In ANNs, neurons are called nodes, and synapses are called edges.
The strength of the synaptic connections is represented by the weight of the edge, and the threshold is determined by the activation function.
An ANN is basically composed of three layers: an input layer, a hidden layer, and an output layer.
A DNN forms a "deep" layered structure by increasing the number of hidden layers and succeeds in understanding the complex structure of the real world.

CNNs are mainly used for image classification.
The input layer of a CNN associates the pixels of input images with nodes and passes them to the hidden layers.
The hidden layers consist of convolutional layers, pooling layers, and fully-connected layers.
The convolutional layers apply small filters to the entire image by sliding them gradually to produce images with a collection of representative values called feature maps.
The fully-connected layers concatenate the features from the convolutional and pooling layers and pass them to the output layer.

\subsection{Fine-tuning}
\label{sec:Fine-tuning}

Fine-tuning is one way of transfer learning~\cite{8107520}.
Transfer learning is a method to apply knowledge gained from one domain to another.
Transfer learning can create highly accurate models even in domains with small amounts of data available by transferring knowledge from domains with large amounts of high-quality data.
Transfer learning can also reduce the model training time by reusing the edge weights of previously trained models.
On the other hand, transfer learning can cause a deterioration in accuracy called negative transfer, which occurs when the transfer method is inappropriate or when the source and destination domains are relatively unrelated.

In transfer learning of CNN, the network trained in one domain is reused for another domain.
In fine-tuning, the basic structure of the CNN network is reused. The edge weights in some layers could be relearned while other layers are frozen, i.e., fixed.
Although relearning is useful for acquiring knowledge in a new domain, relearning the entire model may result in overtraining if the dataset is very small. Overtraining is a problem in which a CNN network closely related to a specific dataset cannot accurately classify additional data.

In short, in order to achieve high classification accuracy through fine-tuning, it is crucial to select the appropriate degree of frozen layers depending on the similarity and size of the dataset to be classified.

\section{Experimental Method}
\label{sec:method}

In this section, we explain the experimental method for this study.
We first explain the CNN models and fine-tuning parameters we used.
Next, we describe how we created the malware image and the characteristics of the dataset we used.
Finally, we describe the evaluation criteria for measuring the accuracy of malware subspecies classification.

\subsection{CNN Models and Fine-tuning Parameters}

We selected 24 models from the ImageNet pre-trained models provided by Keras as our pre-trained models. The models were DenseNet121, DenseNet169, DenseNet201~\cite{Huang_2017_CVPR}, EfficientNetB0, EfficientNetB1, EfficientNetB2, EfficientNetB3, EfficientNetB4, EfficientNetB5~\cite{pmlr-v97-tan19a}, InceptionResNetV2~\cite{AAAI1714806}, InceptionV3~\cite{7780677}, MoblileNet~\cite{howard2017mobilenets}, MobileNetV2~\cite{8578572}, NASNetLarge, NASMobile~\cite{Zoph_2018_CVPR}, ResNet101, ResNet152, ResNet50~\cite{He_2016_CVPR}, ResNet101V2, ResNet152V2, ResNet50V2~\cite{10.1007/978-3-319-46493-0_38}, VGG16, VGG19~\cite{simonyan2014very}, and Xception~\cite{Chollet_2017_CVPR}.

As mentioned in \autoref{sec:Fine-tuning}, it is necessary to pay attention to the degradation of classification accuracy due to negative transfer because the malware images do not seem to resemble pre-trained natural objects. 
Then, since the number of malware appearing in a year depends on the trend of attacks, overtraining should be avoided if the number of data is small.
In order to find the right degree of fine-tuning, it is appropriate to try multiple parameters. we used five different fine-tuning parameters for each model: the ratio of frozen parameters is (1) 1 (Frozen\_all), (2) 3/4 (Frozen\_3/4), (3) 1/2 (Frozen\_1/2), (4) 1/4 (Frozen\_1/4), and (5) 0 (Frozen\_none).
\autoref{tab_targetlayer} shows the number of frozen layers and the name of the border layer for each model.

Note that the number of layers shown in \autoref{tab_targetlayer} is based on the Keras implementation and differs from the number of layers in the theoretical model (e.g., 16 layers in VGG16 or 121 layers in DenseNet121).
In addition, the number of parameters to be frozen is not exactly 1/4, 1/2, or 3/4 of the total number of parameters because the number of parameters in a layer is not constant, and if the model has a block consisting of multiple layers, the layer to be frozen must be specified at the block boundary.

\subsection{Malware Image Conversion Tool}

\begin{table}[t]
	\caption{Rules to determine the image width.}
	\label{tab_imagewidth}
	\centering
	\begin{tabular}{cc}
		\toprule
		File Size Range & Image Width\\
		\midrule
		\textless10KB    & 32 \\
		10KB   -  30KB    & 64   \\
		30KB   -  60KB    & 128  \\
		60KB   -  100KB   & 256  \\
		100KB  -  200KB   & 384  \\
		200KB  -  500KB   & 512  \\
		500KB  -  1,000KB  & 768  \\
		\textgreater=1,000KB & 1024 \\
		\bottomrule
	\end{tabular}
\end{table}

We developed a tool to convert malware files into images. 
This tool is designed to accept files in Portable Executable (PE) format for Windows malware and Dalvik Executable (DEX) format for Android malware.
The tool checks the header information of the file to determine the file format and converts the binary data of the file into a grayscale image with 256 shades per pixel. 
The width of the image is determined by the file size according to the rule in \autoref{tab_imagewidth}.
This rule is equivalent to that proposed by Nataraj et al.~\cite{10.1145/2016904.2016908}, but we extended it to be applied to Android malware. 

The size of the image generated by the malware file varies, but it is automatically changed to the standard size of each model at the time of input by using the nearest-neighbor algorithm, which is the default method for image resizing in Keras (for example, for the VGG16 model, the image size is a 224 x 224 square).

\subsection{Malware Dataset}
\label{sec_Malware_Dataset}

\begin{table}[t]
	\caption{The VirusTotal May 2020 Windows dataset.}
	\label{tab_labeling_windows}
	\centering
	\begin{tabular}{llr}
		\toprule
		Family & Packer / Compiler & Files\\
		\midrule
		Backdoor.Delf    & tElock       & 186\\
		Backdoor.Gobot   & Delphi       & 43 \\
		Backdoor.IRCBot  & Visual C++   & 55\\
		Backdoor.Wabot   & Delphi       & 186\\
		Trojan.Agent     & Visual C++   & 54\\
		Worm.Benjamin    & ASPack       & 92\\
		Worm.Picsys      & UPX          & 126\\
		Worm.Small       & Visual C++   & 79\\
		StormAttack      & Visual C++   & 50\\
		Dropper.Dinwod   & UPX          & 165\\
		Trojan.Agent     & MingWin32    & 66\\
		Trojan.Agent     & Visual Basic & 74 \\
		Trojan.Mansabo   & Visual Basic & 53\\
		Trojan.VB        & Visual Basic & 101\\
		Trojan.Agent.VB  & Visual Basic & 42\\
		AdWare.Gator     & Visual C++   & 57\\
		AdWare.Lollipop  & LCC Win32    & 80\\
		CoinMiner        & Visual C++   & 47\\
		\midrule
		Total & & 1,556\\
		\bottomrule
	\end{tabular}
\end{table}

\begin{table}[t]
	\caption{The VirusTotal May 2020 Android dataset.}
	\label{tab_labeling_android}
	\centering
	\begin{tabular}{lr}
		\toprule
		Family & Files\\
		\midrule
		Android.Agent    & 43\\
		Android.Ewind    & 72\\
		Android.SMSreg   & 49\\
		\midrule
		Total & 164\\
		\bottomrule
	\end{tabular}
\end{table}

We used Malimg and Drebin as the malware datasets to obtain the optimal CNN models for malware image classification to compare the classification accuracy of the previous study.
Malimg is a dataset introduced by Nataraj et al.~\cite{10.1145/2016904.2016908}.
It consists of 9,339 images of malware files in the PE32 format, converted to PNG format images, and labeled with 25 classes using Windows Security Essential.
Drebin is a malware dataset introduced by D. Arp et al.~\cite{arp2014drebin}.
It is an Android malware dataset collected from August 2010 to October 2012, which contains 5,560 APK files of Android malware and is classified into 178 classes based on Kaspersky~\cite{kaspersky-soft}.
To compare with the malware classification experiment of previous research, we selected the top 20 classes from the Drebin dataset.
To make malware images, we obtained the class.dex file from each APK file.
Note that one file was excluded from our dataset because it is a compilation of 25 APK files and we could not obtain a unique class.dex file\footnote{The SHA256 value of this file is:\\
\rightline{df2c357f513c270cd1d06418e4eaf64aeb6b2d947149e83ed4f42c88286b76a7}}, so the total number of malware samples was 4,663.

We also used VirusTotal to validate the classification accuracy for recent malware variants.
VirusTotal offers two types of data access with an academic account: 1) access to the Academic API or 2) access to the malware sample folder.
We accessed the malware sample folder and got the recent malware dataset.
Since new malware data is added to the malware sample folder approximately every six months, we used the datasets added in May 2020 for our evaluation.
From each dataset, we selected the Win32\_EXE category and the Android category for malware classification.
The May 2020 dataset had 38,444 files in the Win32\_EXE category and 434 files in the Android category.

Due to many research reports that malware labels of anti-virus scanners are biased~\cite{10.1145/3463274.3463336,10.1145/3427228.3427261,10.1145/3422337.3447849,8937650,10.1145/3474369.3486873}, we determined to use labels that were matched by at least three scanners for VirusTotal.
However, since VirusTotal uses more than 70 scanners for each malware, it is time-consuming to check all combinations to extract matching labels.
Therefore, we used one scanner to list base labels and then check if two or more scanners had the same labels.
We used Kaspersky for our base labels because it supports both Windows and Android.

One problem with the malware image approach, especially on Windows, is that compilers and packers could drastically change the binary image, even if the source code of the malware is the same.
Therefore, we decided to add the names of the compilers and packers used by the malware to the labels to make the identification easier.
To detect compilers and packers, we used five detectors (PEiD~\cite{PEiD}, DIE~\cite{DIE}, Exeinfo\ PE~\cite{ExeinfoPE}, PE\ Detective~\cite{PEDetective}, and TrID~\cite{TrID}) and used the name of compilers and packers that were matched by at least three detectors.
If the malware uses a compiler or packer that cannot be detected, we expect security analysts to analyze the malware from scratch.

In addition, to reduce the imbalance between malware classes, we selected classes that contained at least 40 malware files and excluded malware with non-family name labels such as ``generic'' and ``gen.''

As a result, we obtained 1,556 Windows malware files in 18 classes from the May 2020 dataset, as shown in \autoref{tab_labeling_windows}, which we call the `VirusTotal May 2020 Windows'' dataset.
For Android, we obtained 164 malware files classified into three classes, as shown in \autoref{tab_labeling_android}, which we call the `VirusTotal May 2020 Android'' dataset.

\subsection{Evaluation Criteria}

Each malware file in the dataset is labeled with the name of the family it belongs to.
Based on the label, we trained our models with the malware images to classify the malware files.
Since there are many types of labels in the dataset, it is a multi-class classification.
Therefore, we evaluated the classification results using the commonly used metrics defined by the following equations~\cite{sokolova2009systematic}: 

\begin{eqnarray*}
  Accuracy &=& \frac{1}{l}\cdot\sum^{l}_{i=1}\frac{tp_{i}+tn_{i}}{tp_{i}+fn_{i}+fp_{i}+tn_{i}} \\
  Precision &=& \frac{1}{l} \cdot \sum^{l}_{i=1} \frac {tp_{i}} {tp_{i}+fp_{i}} \\
  Recall &=& \frac{1}{l} \cdot \sum^{l}_{i=1} \frac {tp_{i}} {tp_{i}+fn_{i}}  \\
  F\mathchar`-score &=& \frac{1}{l} \cdot \sum^{l}_{i=1} \frac {2 \cdot Precision \cdot Recall}{Precision + Recall} \\
\end{eqnarray*}
where $ tp_{i}$ is true positive, 
$fp_{i}$ is false positive,
$fn_{i}$ is false negative, and
$tn_{i}$ is true negative, respectively.
All the metrics use macro averaging.

We evaluated all classifications with stratified 10-fold cross-validation, i.e., for each class, 90\% of the data is used for training and the remaining 10\% for testing. 
The cross-validation involves multiple rounds of creating training and test data in order to evaluate the average. The hold-out validation, on the other hand, does it only once.

\begin{table}[t]
	\caption{Batch size for each model.}
	\label{tab_batchsize}
	\centering
	\begin{tabular}{ll}
		\toprule
		Model & Size \\
		\midrule
		MobileNet, MobileNetV2  & 128 \\  \hline
		DenseNet121, EfficientNetB0, InceptionV3, NASNetMobile   & 64    \\  
		ResNet101, ResNet101V2, ResNet50, ResNet50V2, VGG16        &     \\  
		VGG19 &  \\  \hline
		DenseNet169, DenseNet201, EfficientNetB1, EfficientNetB2 & 32   \\ 
        InceptionResNetV2, ResNet152, ResNet152V2, Xception &      \\  \hline
		EfficientNetB3 & 16   \\  \hline
        EfficientNetB4, NASNetLarge & 8 \\ \hline
		EfficientNetB5 &  4   \\  
		\bottomrule
	\end{tabular}
\end{table}

\section{Evaluation Result}
\label{sec:evaluation}

This section presents our evaluation results.
We first show the classification results of Malimg and Drebin, and then show the results of our attempt to classify malware in the VirusTotal May 2020 dataset.

\subsection{Setup}
\label{sec:setup}

We ran Keras 2.4.3 with Python 3.6.9 on Ubuntu 18.04 LTS 64bit on three machines with Nvidia GeForce RTX 2080Ti. The backend of Keras was TensorFlow 2.2 on CUDA 10.1 and cuDNN 7.6.5. 

We designed the size of the output layer for each model to match the class size of the dataset. 
Specifically, we used \emph{softmax} as the activation function and set the number of outputs to 25 for Malimg, 20 for Drebin, 18 for VirusTotal May 2020 Windows, and 3 for Android, respectively. 
In the fully-connected layer, we used \emph{Relu} as the activation function and set the number of outputs to 128, which was determined by referring to a previous study that used a dataset with a similar number of classes and samples~\cite{10.1145/3342999.3343002}. 

The hyperparameters of our fine-tuned models were as follows: the loss function was categorical cross-entropy, the optimizer was SGD, the learning rate was 1e-4, the momentum was 0.9, and the epochs numbered 30. 
The batch size for each model was determined to be the value shown in \autoref{tab_batchsize}, which was the highest possible value to shorten the learning time.

Data augmentation, which increases the number of images by flipping, rotating, scaling, etc., is generally effective in obtaining high classification accuracy from a limited data set in machine learning.
However, in our preliminary experiments, the classification accuracy with data augmentation was lower than that without augmentation, so we did not use it this time.

\subsection{Classification of Malimg}
\label{sec_malimg}

\begin{figure}[t]
	\centering
	\includegraphics[width=\linewidth,clip]{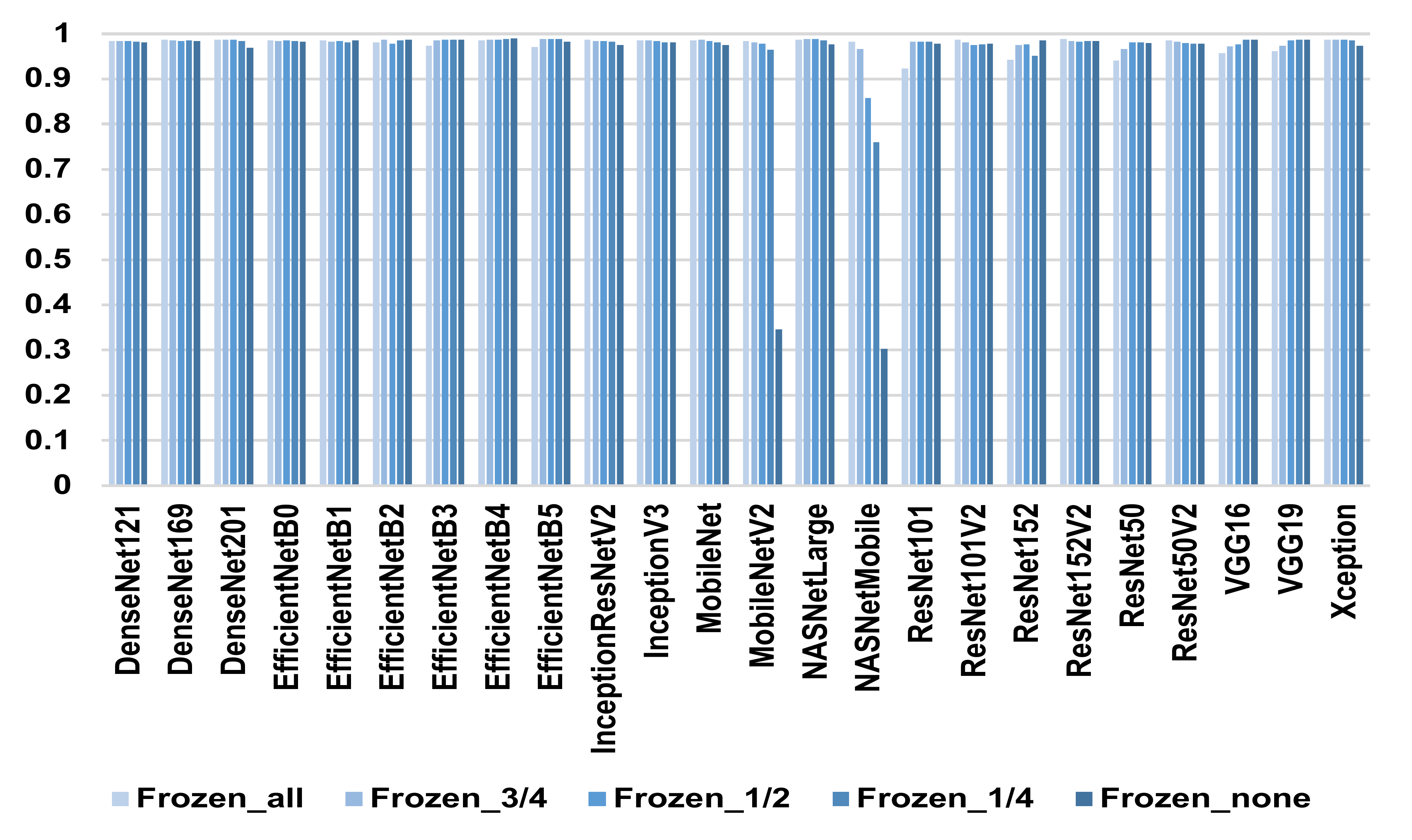}
	\caption{Classification accuracy of Malimg.}
	\label{fig_malimg}
\end{figure}

\begin{table}[t]
	\caption{Top five models for Malimg.}
	\label{tab_top5_malimg}
	\centering
	\scalebox{0.95}{
	\begin{tabular}{cccccc} 
		\toprule
	    Model & Frozen & $Accuracy$ & $Precision$ & $Recall$ & $F\mathchar`-score$\\
		\midrule
		EfficientNetB4 &none & 0.9896 & 0.9739 & 0.9718 & 0.9714\\ 
		EfficientNetB5 &1/2  & 0.9882 & 0.9704 & 0.9685 & 0.9676\\
		EfficientNetB5 &3/4  & 0.9881 & 0.9732 & 0.9681 & 0.9663\\
		NASNetLarge    &3/4  & 0.9877 & 0.9687 & 0.9666 & 0.9662\\
		EfficientNetB5 &1/4  & 0.9877 & 0.9686 & 0.9662 & 0.9662\\
		\bottomrule
	\end{tabular}
	}
\end{table}

\begin{figure}[t]
	\centering
	\includegraphics[width=\linewidth,clip]{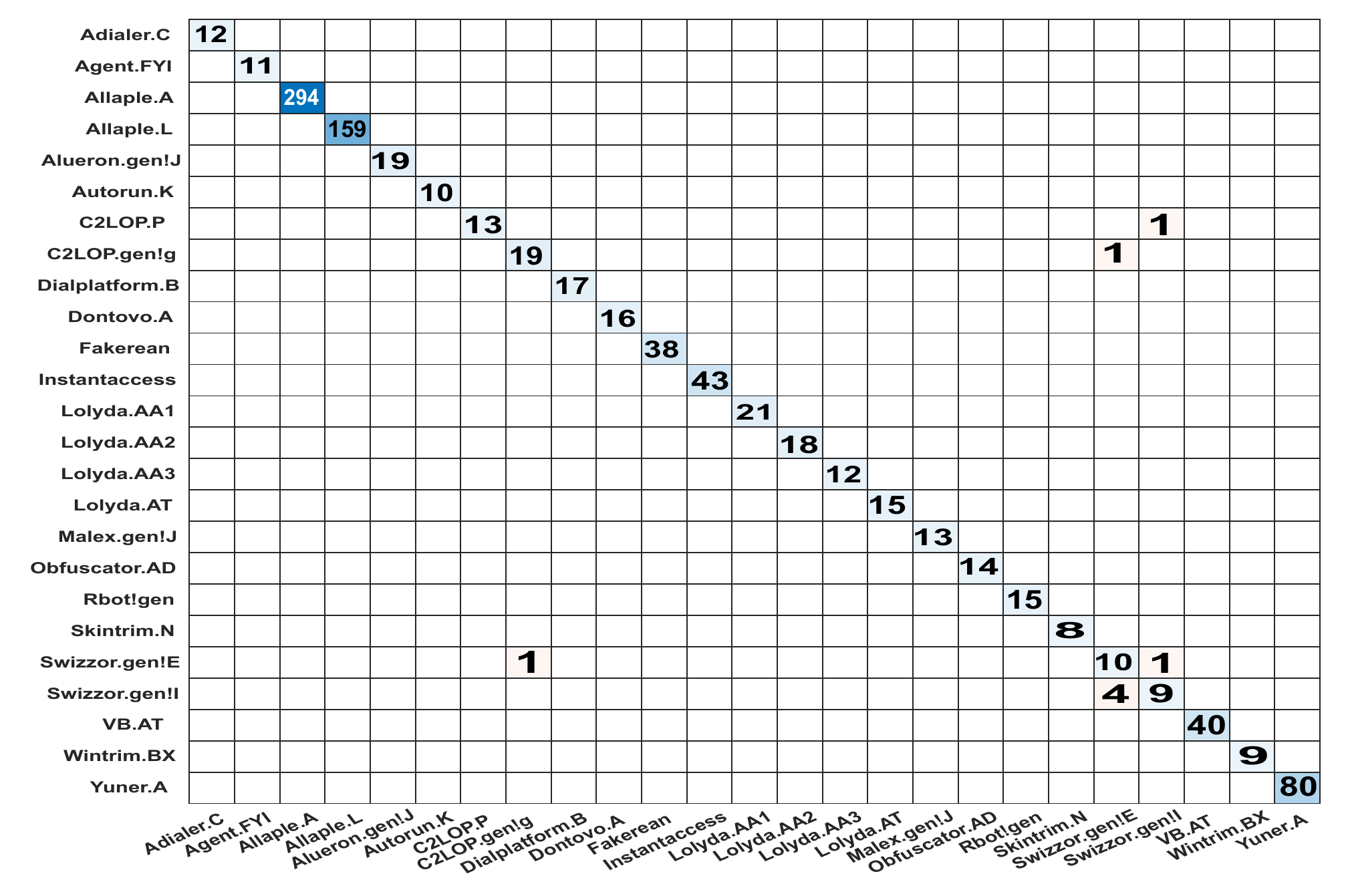}
	\caption{Confusion matrix of EfficientNetB4 Frozen\_none for Malimg.}
	\label{fig_malimg_cm}
\end{figure}

\begin{table}[t]
	\caption{Comparison of cross-validation for Malimg.}
	\label{tab_comparison_malimg_cross}
	\centering
	\begin{tabular}{ccc}
		\toprule
		Method & $Accuracy$ & $F\mathchar`-score$\\
		\midrule
CNN \cite{8330042} & 94.50\% & N/A  \\
GIST + K-nearest neighbors  \cite{10.1145/2016904.2016908} & 97.18\% & N/A \\
ResNet50 \cite{8260773} & 98.62\% & N/A \\
EfficientNetB4 + Frozen\_none (ours)& \textbf{\underline{98.96\%}} & \textbf{\underline{97.14}}\% \\
		\bottomrule
	\end{tabular}
\end{table}

\begin{table}[t!]
	\caption{Comparison of hold-out validation for Malimg.}
	\label{tab_comparison_malimg_hold}
	\centering
	\begin{tabular}{ccc}
		\toprule
		Method & $Accuracy$ & $F\mathchar`-score$\\
		\midrule
VGG16 \cite{10.1145/3320326.3320333} & 97.02\% & N/A \\
M-CNN \cite{8328749} & 98.52\% & N/A\\
IMCFN \cite{VASAN2020107138} & 98.82\% & 98.75\% \\
Xception \cite{8763852} & 99.03\% & N/A \\
EfficientNetB4 + Frozen\_none (ours)& \textbf{\underline{99.13\%}} & 97.66\%  \\
IMCEC \cite{VASAN2020101748} & 99.50\% & 99.48\% \\
		\bottomrule
	\end{tabular}
\end{table}

We first classified Malimg.
\autoref{fig_malimg} shows the classification accuracy of the stratified 10-fold cross-validation for each model.
Most of the models except for the MobileNetV2 and NASNetMobile achieved over 90\% accuracy. Some models of the MobileNetV2 and NASNetMobile were below 40\%.
As the name implies, MobileNetV2 and NASNetMobile are designed to work with limited system resources and have a smaller number of training parameters than the other models. For example, the VGG16 without the fully-connected layers has 14,714,688 training parameters, while the MobileNetV2 without them has 2,257,984 and the NASNetMobile without them has 4,269,716. 
We consider that the small number of training parameters may have affected the classification accuracy.

\autoref{tab_top5_malimg} shows the top five models in classification accuracy. 
The model with the highest accuracy was EfficientNetB4 Frozen\_none, with a classification accuracy of 98.96\%.
Looking at the results of the ten tests among the cross-validation using the EfficientNetB4 Frozen\_none, the maximum accuracy was 0.9913, the minimum accuracy was 0.9869, and the standard deviation was 0.0017. 
\autoref{fig_malimg_cm} shows the confusion matrix when the maximum accuracy was obtained. The number of training data was 8,416, the number of test data was 923, $accuracy$ was 0.9913, $recall$ was 0.9761, $precision$ was 0.9773, and $F\mathchar`-score$ was 0.9766. Malimg is an imbalanced dataset; for example, the Allaple.A family has 294 test data, while the Skintrim.N family has only eight test data. 
In general, when classifying imbalanced data sets, correctly classifying classes with more data may lead to higher accuracy, even if classes with fewer data are not correctly classified. However, in this classification, there was no class with extremely poor accuracy. 
This indicates that our model correctly extracted the features of each class.

As shown in \autoref{tab_comparison_malimg_cross} (cross-validation), our method obtained the highest $accuracy$ and $F\mathchar`-score$. 
Achieving high classification $accuracy$ and a high $F\mathchar`-score$ is very important in terms of reducing the burden on security analysts. 
As shown in \autoref{tab_comparison_malimg_hold} (hold-out validation), our method was the second best.
Comparing cross-validation and hold-out validation, cross-validation is more important during evaluation. This is because an accuracy obtained by hold-out validation may decrease substantially depending on how the training and test data are created.

\subsection{Classification of Drebin}
\label{sec_drebin}

\begin{figure}[t]
	\centering
	\includegraphics[width=\linewidth,clip]{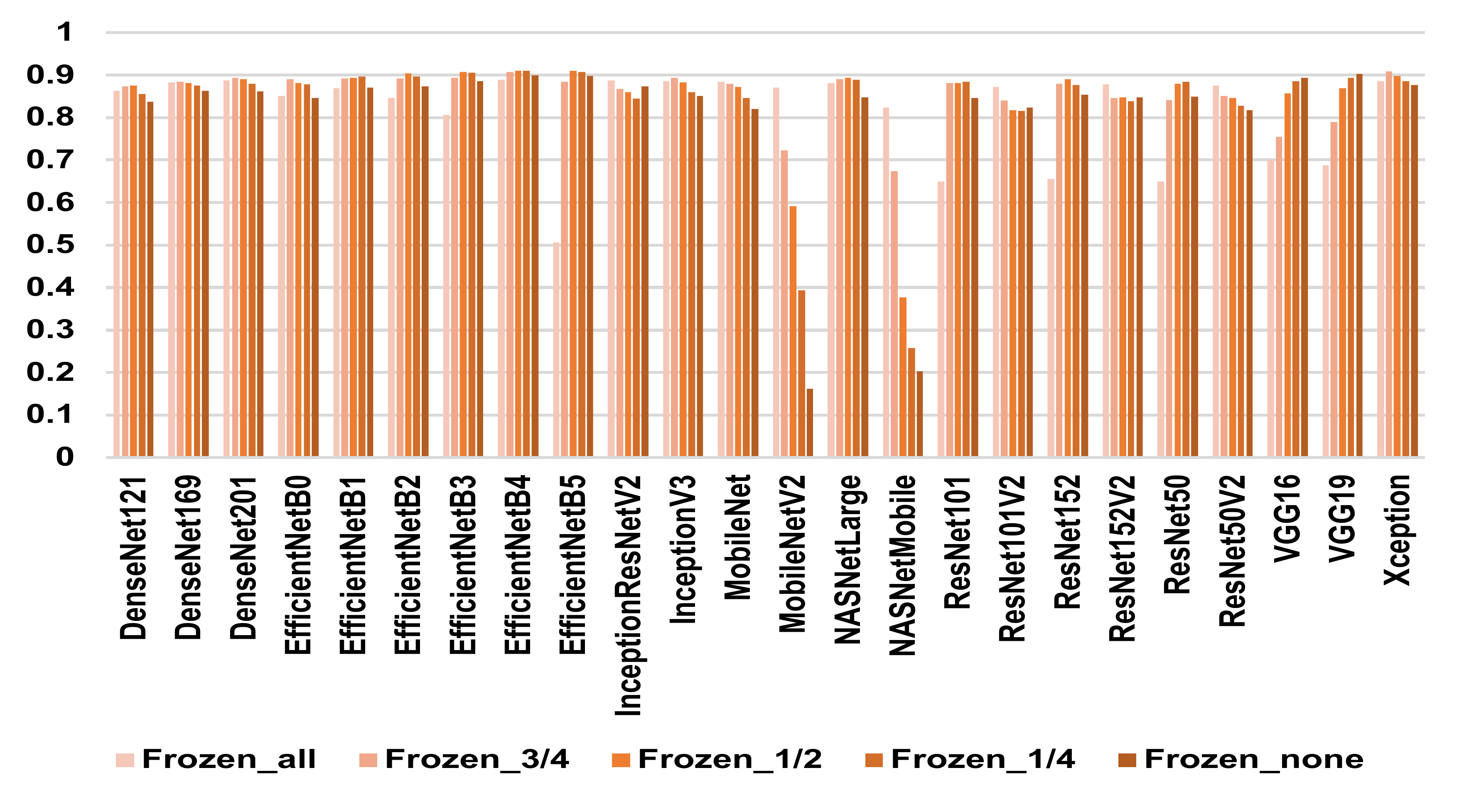}
	\caption{Classification accuracy of Drebin.}
	\label{fig_drebin}
\end{figure}

\begin{table}[t]
	\caption{Top 5 models for Drebin.}
	\label{tab_top5_drebin}
	\centering
	\scalebox{0.95}{
	\begin{tabular}{cccccc} 
		\toprule
	    Model & Frozen & $Accuracy$ & $Precision$ & $Recall$ & $F\mathchar`-score$\\
		\midrule
EfficientNetB4 & 1/4 & 0.9103 & 0.8707 & 0.8524 & 0.8536\\
EfficientNetB5 & 1/2 & 0.9098 & 0.8777 & 0.8458 & 0.8517\\
EfficientNetB4 & 1/2 & 0.9098 & 0.8732 & 0.8449 & 0.8498\\
EfficientNetB5 & 1/4 & 0.9079 & 0.8847 & 0.8371 & 0.8490\\
EfficientNetB3 & 1/2 & 0.9071 & 0.8645 & 0.8365 & 0.8404\\
		\bottomrule
	\end{tabular}
	}
\end{table}

\begin{figure}[t]
	\centering
	\includegraphics[width=\linewidth,clip]{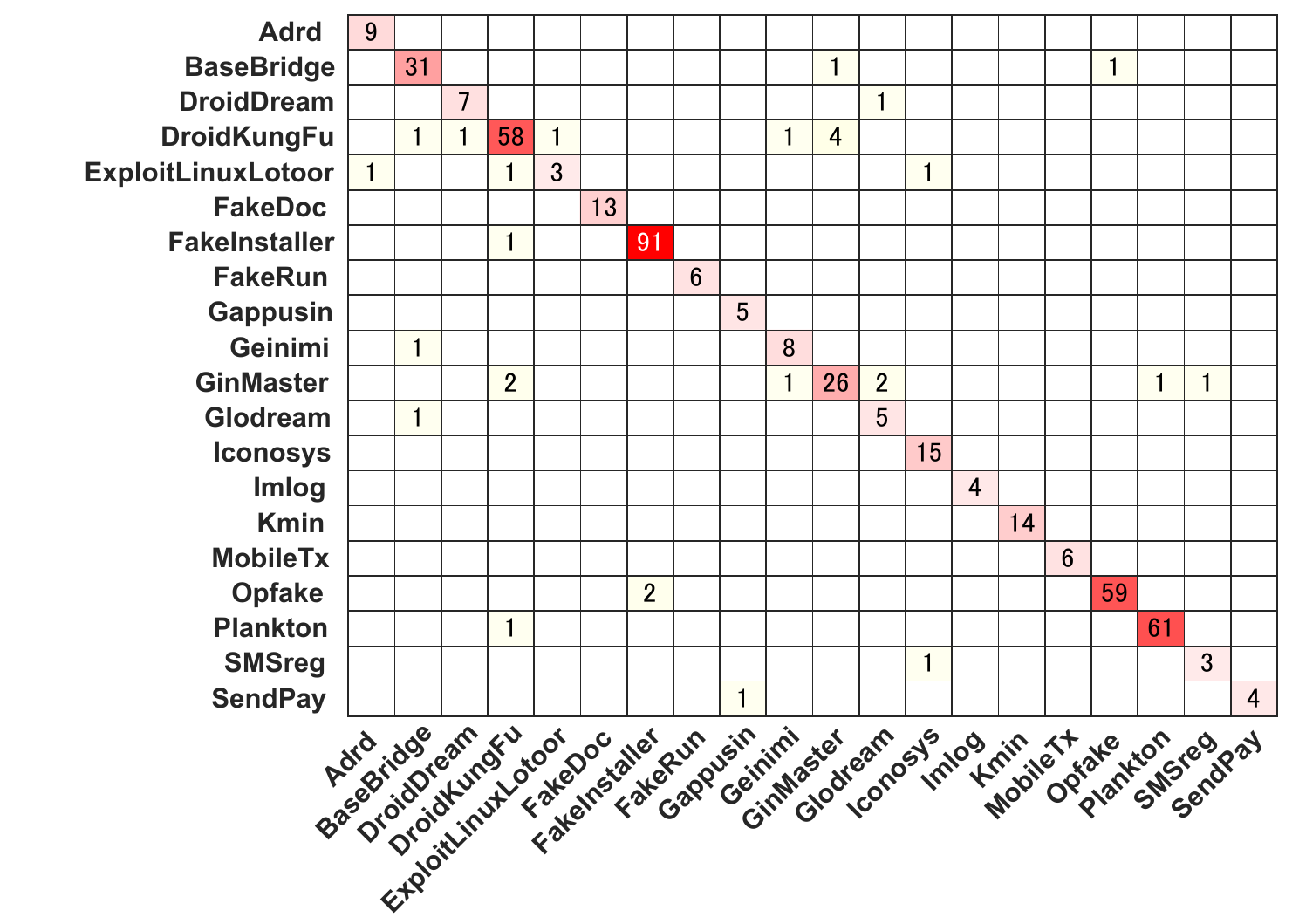}
	\caption{Confusion matrix of EfficientNetB4 Frozen\_1/4 in Drebin.}
	\label{fig_drebin_cm}
\end{figure}

Next, we classified Drebin. \autoref{fig_drebin} shows the classification accuracy on closs-validation. 
Similar to Malimg, we found MobileNetV2 and NASNetMobile to be less accurate.
\autoref{tab_top5_drebin} shows the top five models in classification accuracy. 
The model with the highest accuracy was the EfficientNetB4 Frozen\_1/4 model with a classification accuracy of 91.03\%.
Looking at the results of the ten tests among the cross-validation using EfficientNetB4 Frozen\_1/4 model, the maximum accuracy was 0.9365, the minimum accuracy was 0.8781, and the standard deviation was 0.014. 
\autoref{fig_drebin_cm} shows the confusion matrix when the maximum accuracy was obtained. The number of training data was 4,206, the number of test data was 457, $accuracy$ was 0.9365, $recall$ was 0.9096, $precision$ was 0.9016, and $F\mathchar`-score$ was 0.9055. 
There was no class with extremely poor classification accuracy, but the classification accuracy was not sufficient for classes with little data, such as Exploit Linux Lotoor, SMSreg and SendPay.

\begin{table}[t]
	\caption{Comparison of hold-out validation for image-based Drebin.}
	\label{tab_comparison_drebin}
	\centering
	\begin{tabular}{ccc}
		\toprule
		Method & $Accuracy$ & $F\mathchar`-score$\\
		\midrule
		Feature Fusion-SVM \cite{9461194} & 93.24\% & N/A\\
		EfficientNetB4 + Frozen\_1/4 (ours) & \textbf{\underline{93.65\%}} & \textbf{\underline{90.55\%}} \\
		\bottomrule
	\end{tabular}
\end{table}

To the best of our knowledge, no previous study has evaluated the classification accuracy of image-based Drebin with cross-validation. 
As shown in \autoref{tab_comparison_drebin}, our classification accuracy was higher than that of the previous study using image-based Drebin with hold-out validation. 
Since our method only converts the dex file contained in the APK file into an image, the pre-processing is very simple. Then, the features extracted from images work effectively to classify malware variants.

\subsection{Windows Malware in VirusTotal}
\label{sec_windows}

To examine the classification accuracy of recent real Windows malware, we used the VirusTotal May 2020 Windows dataset and the models that had high classification accuracy for Malimg.

\begin{figure}[t]
	\centering
	\includegraphics[width=\linewidth,clip]{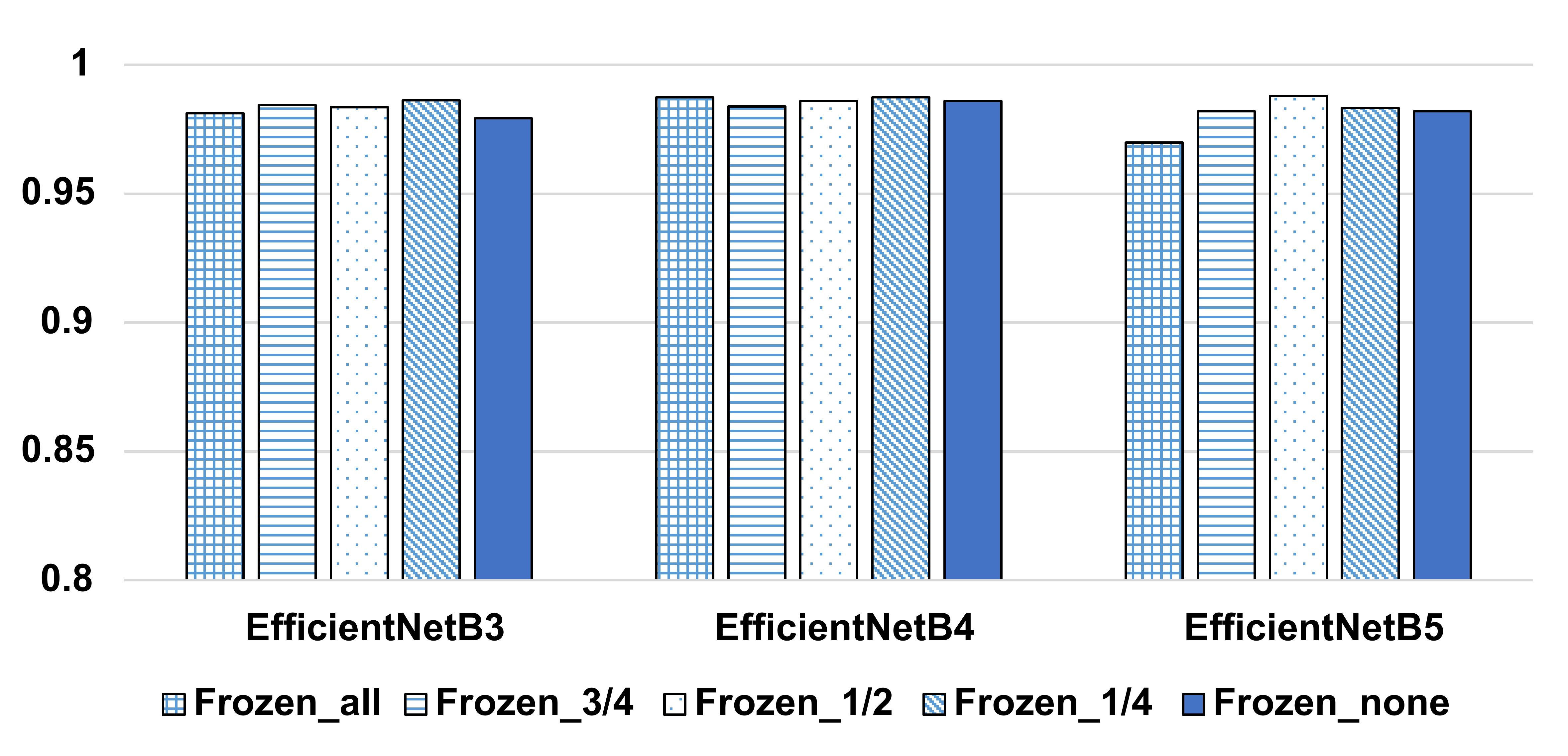}
	\caption{Classification accuracy for VirusTotal May 2020 Windows.}
	\label{fig_windows}
\end{figure}

\begin{table}[t]
  \caption{Top five models for VirusTotal May 2020 Windows.}
  \label{tab_top5_windows}
  \centering
	\scalebox{0.95}{
	\begin{tabular}{cccccc}
    \toprule
	    Model & Frozen & $Accuracy$ & $Precision$ & $Recall$ & $F\mathchar`-score$\\
    \midrule
    EfficientNetB5 & 1/2  & 0.9878 & 0.9868 & 0.9858 & 0.9844\\
    EfficientNetB4 & 1/4  & 0.9874 & 0.9880 & 0.9849 & 0.9851\\
    EfficientNetB4 & all  & 0.9871 & 0.9893 & 0.9849 & 0.9859 \\
    EfficientNetB3 & 1/4  & 0.9862 & 0.9887 & 0.9837 & 0.9849 \\
    EfficientNetB4 & 1/2  & 0.9858 & 0.9888 & 0.9838 & 0.9846 \\
   \bottomrule
\end{tabular}
	}
\end{table}

\autoref{fig_windows} shows the classification accuracy of the VirusTotal May 2020 Windows dataset for each model. 
We selected EfficientNetB3, B4, and B5 with five fine-tuning parameters. 
Most of the models had an accuracy of 96\% or better, and no model had a lower accuracy.
\autoref{tab_top5_windows} shows the top five models with the highest classification accuracy. 
As shown in the table, the EfficientNetB5 Frozen\_1/2 model showed the highest classification accuracy of 98.78\%, which was slightly lower than Malimg's best classification accuracy (98.96\%). 
However, the difference between $Accuracy$ and $F\mathchar`-score$ was 0.0034, which was better than 0.0182 in the best model in Malimg.

\begin{figure}[t]
  \centering
  \includegraphics[width=\linewidth,clip]{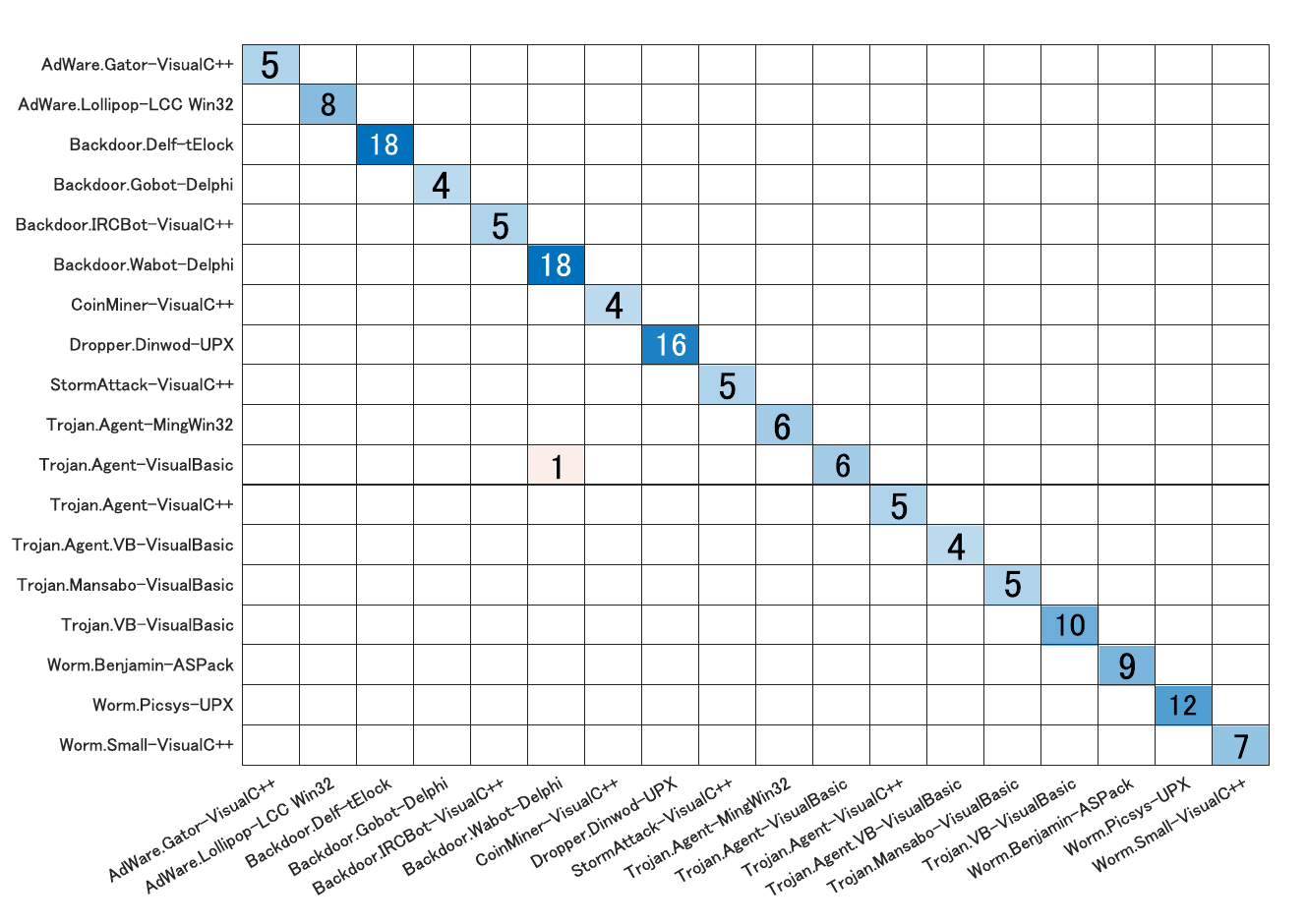}
  \caption{Confusion matrix of EfficientNetB5 Frozen\_1/2 in VirusTotal May 2020 Windows.}
  \label{fig_windows_cm}
\end{figure}

Looking at the results of the ten tests among the cross-validation using the EfficientNetB5 Frozen\_1/2 model, the maximum accuracy was 0.9932, the minimum accuracy was 0.9662, and the standard deviation was 0.0084. 
\autoref{fig_windows_cm} shows the confusion matrix at maximum accuracy. 
The number of training data was 1,408 and test data was 148, $accuracy$ was 0.99324, $recall$ was 0.9920, $precision$ was 0.9970, and $F\mathchar`-score$ was 0.9942. Fortunately, in the confusion matrix, almost all the classes were fully classified. 
Overall, our method correctly classified most of the VirusTotal May 2020 Windows dataset. 

\subsection{Android Malware in VirusTotal}

\begin{figure}[t]
  \centering
  \includegraphics[width=\linewidth,clip]{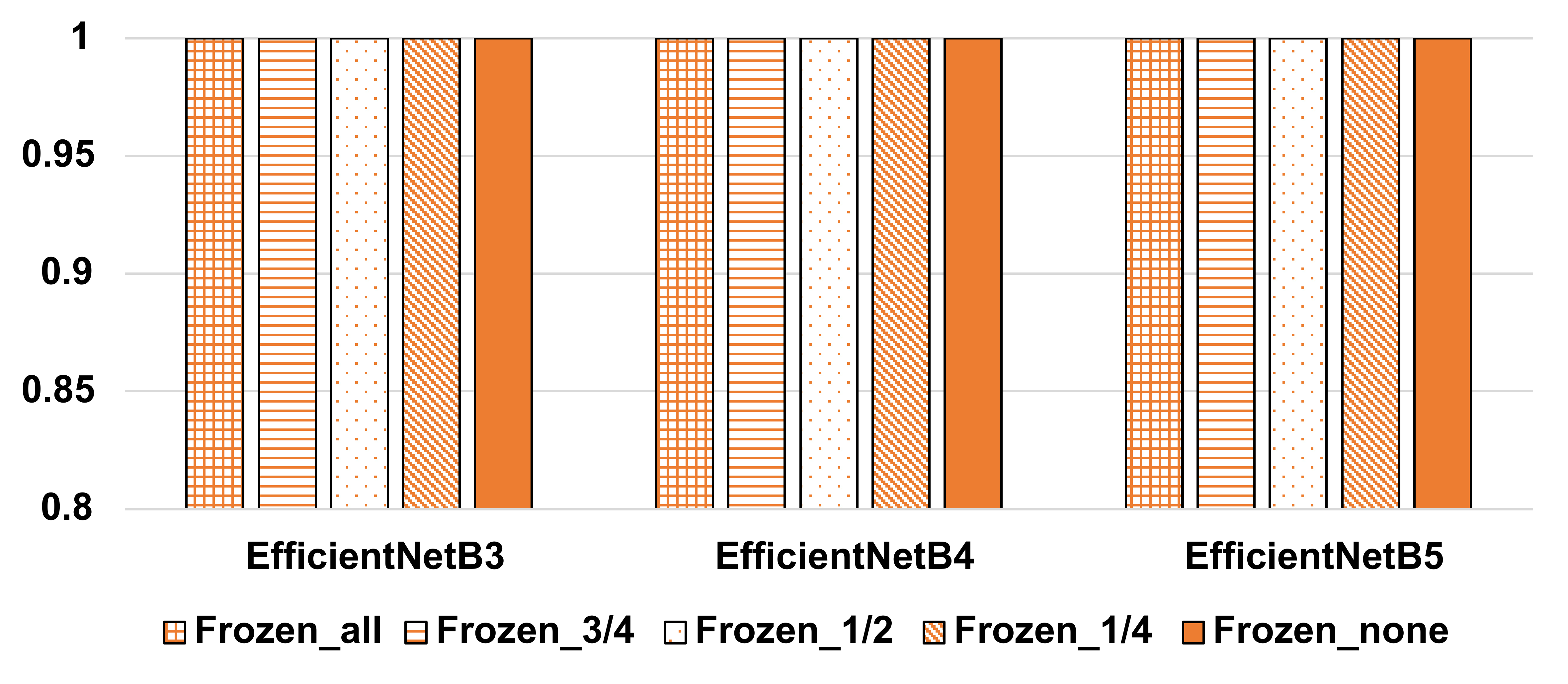}
  \caption{Classification accuracy for VirusTotal May 2020 Android.}
  \label{fig_Android}
\end{figure}

Finally, we performed the same evaluation for the Android malware.
We used EfficientNetB3, B4, and B5 to classify the Virus Total May 2020 Android dataset. 
\autoref{fig_Android} shows the classification accuracy of each model. The results show that the classification accuracy of all models reached 100\%. 

Since the Android dataset had only three classes, there was no difference in the classification accuracy even with different models and degrees of fine-tuning, respectively.
In addition, the number of files in the Android dataset was 146, which was less than the other datasets (Malimg: 9,339, Drebin: 4663, VirusTotal May 2020 Windows: 1,556), but the classification accuracy by cross-validation reached 100\%. It indicates no overtraining in EfficientNetB3-B5, even when all the parameters are not frozen.

\subsection{Discussion and Future Work}

We discuss some of the issues regarding our evaluation results and present some future work.
First of all, the evaluation result shows that EfficientNetB4 fine-tuned by freezing no or only 1/4 of the pre-trained parameters, had the highest classification accuracy on the Malimg and Drebin datasets.
As for the VirusTotal Windows dataset, we found that EffieicnetNetB5 fine-tuned with 1/2 frozen achieved the highest classification accuracy for the May 2020 Windows dataset, while the EfficientNetB3-B5 with no fine-tuning achieved 100\% accuracy for the May 2020 Android dataset.

These results suggest that the optimal model for malware classification was obtained by using the latest models and reducing the degree of transfer learning.
By reducing the degree of transfer learning, it has become possible to use more layers to learn the features of malware images, which is thought to have helped improve the classification accuracy.
This is due to the fact that with the continued development of malware, it has become possible to extract the characteristics from a large number of malware images.
On the other hand, the classification of Android malware in VirusTotal is considered to be highly accurate regardless of the degree of transfer learning because the number of classes to be classified is small for the period of May 2020.

Based on the experimental results, we expect that the degree of transfer learning most effective in classifying malware is in the range of 0-1/2. 
To search for the optimal deep learning models, we believe that fine-tuning the latest models while gradually lowering the number of frozen parameters from 50\% would be effective as a practical approach.

Secondly, in order to address the problem of Windows malware being widely obfuscated, we added the names of the compiler or packer used by the malware to the labels.
In \autoref{sec_windows}, we have distinguished recent malware variants with high accuracy.
Experience shows that even obfuscated malware may still retain some similarity to the original malware if the same packer and compiler are used.
However, there are many types of obfuscation techniques, and their number is increasing. Malware variants that apply various obfuscation techniques to the original are hard to classify into the same family because their malware image has different characteristics, which is a limitation of image-based malware classification.
Improving resistance to obfuscation and the specific evaluation are future works.

Thirdly, we developed an image conversion tool that supports Windows PE and Android DEX format.
The multi-platform tool has worked well in the evaluations and is expected to help reduce the burden on security analysts.
Nevertheless, some malware runs on other platforms, such as iOS, macOS, and Linux.
We believe our tool can be extended for other platforms and confirm its effectiveness in the future.

Finally, if an attacker modifies a large part of existing malware or adds many new features, the malware would no longer be regarded as a variant.
Therefore, it is necessary to update deep learning models with knowledge of new malware regularly.
Unfortunately, since regular updates typically take days to weeks, we have a plan to introduce some complementary applications to detect malware that emerges in the meantime.

\section{Conclusion}
\label{sec:conclusion}

We conducted an exhaustive study on the impact of deep learning models and the degree of transfer learning on the accuracy of image-based malware classification using 120 combinations with 24 different ImageNet pre-trained CNN models and five levels of fine-tuning parameters.
As a result, we found that the EfficientNetB4 model fine-tuned by freezing no or only 1/4 of the pre-trained parameters had the highest classification accuracy for the Malimg (98.96\%) and Drebin (93.65\%) datasets.
These are the highest classification accuracies ever known to have been validated in cross-validation. 

As for recent real malware, we found that the EffieicnetNetB5 model fine-tuned with 1/2 frozen parameters achieved the highest classification accuracy of 98.78\% for the May 2020 Windows dataset, while the EfficientNetB3 to B5 models with no frozen fine-tuning parameters achieved the 100\% accuracy for the VirusTotal Android dataset.

The experimental results show that the classification accuracy of malware variants tends to be the highest when using the latest deep learning models with a relatively low degree of transfer learning.
A practical approach for exploring the optimal model would be fine-tuning the latest modes while gradually reducing the number of frozen parameters from half.

Future work includes addressing program obfuscation, multi-platform development, and knowledge updating.

\bibliographystyle{IEEEtran}
\bibliography{IEEEabrv,paper}

\end{document}